\documentstyle[12pt]{article}
\begin{document}
\input amssym.def
\input amssym
\font\eightrm=cmr8
\font\twelverm=cmr12

\title{Remarks on unsharp quantum observables,
objectification, and modal interpretations\thanks{
In: Proceedings, The Modal Interpretation of Quantum
Mechanics, Utrecht, 12-14 June 1996. Eds.\ D.\ Dieks,
P.\ Vermaas, Elsevier, to appear 1997.}}
 
\author{Paul Busch\\
{\rm Department of Mathematics,
The University of Hull, UK}\\
{\rm e-mail: p.busch@maths.hull.ac.uk}
}
\date{\eightrm 30 October 1996; Revised: 25 April 1997 and 4
September 1997.}
\maketitle


\section{Introduction}

In this contribution I wish to pose a question that emerged
from discussions with Dennis Dieks regarding my talk at the
Utrecht Workshop. The question is whether a variant of a
modal interpretation is conceivable that could accommodate
property ascriptions associated with nonorthogonal
resolutions of the unity and nonorthogonal families of
relative states as they occur in imperfect or genuinely
unsharp measurements. To explain this question I will review
a recent formulation of the quantum measurement problem in
the form of an insolubility theorem that incorporates the
case of unsharp object observables as well as certain types
of unsharp pointers.  In addition to demonstrating the
necessity for some modification of quantum mechanics, this
allows me to specify the logical position of the modal
interpretations as a resolution to the measurement problem
and to indicate why I think their current versions are not
yet capable of dealing adequately with unsharp quantum
observables. The technical tools that will have been
explained along this line of reasoning will finally serve to
make precise the notion of (unsharp) value ascription that I
would find desirable for a modal interpretation to
ascertain. I will begin with summarising the essence of
unsharp quantum observables, pointing out their ubiquity and
inevitability. In turn I hope to indicate how unsharp observables
might contribute to avoiding some of the difficulties
encountered in the
present modal interpretations.

\section{Unsharp quantum observables}

The starting point for the present considerations is the
fact that measurable quantities are to be represented in
quantum mechanics as {\sl positive operator valued} (POV)
{\sl measures}, which include the more familiar spectral
measures and the associated self-adjoint operators as
special cases corresponding to fairly idealised situations.
The motivations for this extended notion of observables have
been reviewed in great detail in \cite{oqp}. Suffice it to
recall here that a POV measure arises naturally as the
representation of the totality of statistics of measurement
outcomes for all states of the measured system. The idea
underlying this characterisation of an observable is this:
given an observable, the quantum formalism assigns to every
state a probability measure describing the outcome
statistics for that observable, if measured on that
state. Conversely, if the statistics of a given type of
measurement are known for a sufficient number of states
(ideally for all states), then it will be possible to infer
uniquely the observable that has been measured.

This minimal notion of an observable -- and of a measurement -- 
is captured in the so-called {\sl probability
reproducibility} condition. The essential elements of a
measurement are conveniently summarised in the concept of a
{\sl measurement scheme}, represented as a quadruple
${\cal M}:=\langle{\cal H}_{\cal A}, \rho_{\cal A}, U,Z\rangle$,
where ${\cal H}_{\cal A}$ denotes the Hilbert space of the 
measuring device (or probe) ${\cal A}$, $Z$ the pointer
observable of $\cal A$,
i.e., a POV measure on some measurable space
$(\Omega,\Sigma)$, $\rho_{\cal A}$ a fixed initial state of
$\cal A$, and $U$ the unitary measurement coupling serving
to establish a correlation between the object system $\cal
S$ (with Hilbert space $\cal H$) and $\cal A$. Any
measurement scheme $\cal M$ fixes a unique observable of
$\cal S$, that is, a POV measure $E$ on $(\Omega,\Sigma)$
such that the following condition is fulfilled:
\begin{itemize}
\item
 probability reproducibility condition:
$$
{\rm tr}[I\otimes Z(X)\,U\rho\otimes\rho_{\cal A}U^*]
\ =\ {\rm tr}[E(X)\rho]\eqno{\rm (PR)}
$$
for all states $\rho$ of $\cal S$ and all outcome sets
$X\in\Sigma$. 
\end{itemize}
$E$ is the observable measured by means of 
$\cal M$. Conversely, if an observable $E$ of $\cal S$ is
given, then this condition determines which measurement
schemes $\cal M$  serve as measurements of $E$.

Given an observable $E$ on $\big(\Bbb R,{\cal B}(\Bbb
R)\big)$ [where ${\cal B}(\Bbb R)$ is the real Borel algebra] and a
state $\rho$, one can compute the moments of the statistics,
$E^{(n)}_\rho:= \int x^n\,{\rm tr}[\rho E(dx)]$ (when they
exist), and these may lead to the definition of the moment
operators $E^{(n)}:=\int x^nE(dx)$ (provided that suitable
domains do exist). These operators will be symmetric but not
in general self-adjoint. If $E$ is not a PV measure, then
one will {\sl not} have the relation
$E^{(n)}={E^{(1)}}^n$. This shows that a general POV measure
cannot be represented in general by a single operator but
only (if at all) by an infinity of operators whose
expectations would give the moments and thus reproduce the
statistics. In other words, the representation of an
observable by means of a single (self-adjoint) operator is a
rather accidental exceptional case, and in the general case
the full information available in a measurement is to be
encoded in a POV measure.

Deviations from projection valued (PV) measures occur for
two distinct kinds of reasons. First, POV measures may
account for the ever-present imperfections of any
measurement. Second there are measurement situations for
which there exists no ideal background observable
representable as a PV measure.  Accordingly, POV measures
which are not PV measures are referred to as {\sl unsharp
observables} while PV measures correspond to {\sl sharp
observables}. As an example we may mention that there exists
a class of POV measures that can be considered as genuine
phase space observables in every physically relevant
respect. In particular they are covariant under the
Heisenberg group, a feature that is lacking, for instance,
in von Neumann's construction (Section V.4 of \cite{neu}) of
an approximate joint observable for position and momentum
which is based on sharp observables. Such phase space
observables are genuinely unsharp, a reflection of the fact
that the individual measurement outcomes are {\sl fuzzy}
phase space points in accordance with the Heisenberg
uncertainty relation. Moreover, the phase space POV measures
seem to provide the appropriate tool for describing
macroscopic pointers and their quasi-classical measurability
(cf.\ Chapter VI of \cite{oqp}). It has been argued on
general consistency grounds that macroscopic observables
displaying permanent definite values are to be described by
POV measures that are not PV measures \cite{lud}. More 
precisely, the notion of quantities having definite values
at all times, as it is described in classical theories
typically applying to macroscopic objects, can only be
represented approximately within a quantum mechanical
many-body theory, and the quantum representatives of such
quantities are non-PV POV measures.

\section{The objectification problem}

The probability reproducibility condition specifies what it
means that a measurement scheme serves to measure a certain
observable. However, this condition does not exhaust the
notion of measurement. In fact the reproduction of
probabilities in the pointer statistics requires first of
all that in each run of a measurement a pointer reading will
occur; in other words: it is taken to be part of the notion
of measurement that measurements do have definite outcomes.
While the concept of a measurement scheme allows one to {\sl
describe} what happens to the object and apparatus when an
outcome arises, quantum mechanics is facing severe
difficulties to {\sl explain} the occurrence of such
outcomes. This problem arises if one starts with the
interpretational idea that an observable has a definite
value when the object system in question is in an eigenstate
of that observable. If a probe system is coupled to that
object, then probability reproducibility requires that the
corresponding value is indicated with certainty by the
pointer reading after the measurement interaction has
ceased. In this way a definite value of the object
observable leads deterministically to a definite value of
the pointer observable. However, if the object is {\sl not}
in an eigenstate, the observable cannot be ascertained to
have a definite value, and by the linearity of the unitary
measurement coupling, the compound object plus probe system
ends up in a state in which it cannot be ascertained, by
virtue of the eigenstate-eigenvalue rule, that the pointer
has a definite (sharp) value. This is the measurement
problem, or the problem of the objectification of pointer
values. 

Resolutions to this problem are being sought by changing the
rules of the game: either on the side of the formalism
-- introduction of classical observables, or modified
dynamic --, or on the interpretational side -- hidden variables
theories such as `Bohmian mechanics', or various `no-collapse'
interpretations. Before embarking on such radical
revisional programmes, it seems fair to make sure that the 
measurement problem is not merely a consequence of overly
idealised assumptions that would disappear in a more realistic
account. It turns out, however, that the problem does
persist even when measurements are allowed to be inaccurate
and the measuring system is in a mixed rather than a pure
state. The development of these arguments is reviewed in 
\cite{insol}, where an insolubility theorem is given that
pertains to measurements of sharp and unsharp object
observables. This result has recently been overtaken 
by H. Stein who showed that the objectification problem
persists for arbitrary measurement schemes even when the
measurement is not applicable to all
object preparations but only to states in some subspace 
\cite{stein}. A final step can be
taken if the pointer observable is also allowed to be
unsharp. As long as the pointers are still such that they
will assume definite sharp values, the statement of the
insolubility theorem still holds true \cite{uob}. 

In order to give the precise statement of the insolubility
theorem, let us consider a measurement scheme $\cal M$.  The
theorem is based on the following requirements as 
necessary conditions for the objectification of sharp values
of the pointer $Z$ in the postmeasurement state
$\rho'_{\cal SA}\equiv U\,\rho_{\cal S}\otimes\rho_{\cal A}\,U^*$: 

\begin{itemize}
\item {\sl Pointer mixture condition}:
$$
\rho'_{\cal SA}\ =\sum_i I\otimes Z(X_i)^{1/2}\rho'_{\cal SA}
\,I\otimes Z(X_i)^{1/2}\ \equiv\ \sum_i \rho'_{\cal SA}(X_i)
\eqno{\rm (PM)}
$$
for some partition $\Omega=\cup X_i$ and all initial object
states $\rho$;
\item {\sl Pointer value definiteness}:
$$
{\rm tr}\big[I\otimes Z(X_i)\,\rho'_{\cal SA}(X_i)\big]
\ =\ {\rm tr}\big[\rho'_{\cal SA}(X_i)\big] 
\eqno{\rm (PVD)}
$$
for all $i$ and all initial object
states $\rho$.
\end{itemize}
For a derivation of these conditions, see \cite{qtm}.
The first says that the postmeasurement state should be a
mixture of pointer eigenstates, while the second requires
that the final states conditional on reading a result in
$X_i$ are indeed eigenstates of the pointer for which $X_i$
has probability one to occur again upon immediate repetition
of the reading of the pointer observable $Z$. 

The insolubility theorem states: if a measurement scheme 
$\cal M$ fulfills (PM) and (PVD), then the measured observable $E$
according to (PR) is {\sl trivial}; that is,
$E(X)=\lambda(X)I$ for all $X\in\Sigma$, where $\lambda$ is
a state-independent probability measure on
$(\Omega,\Sigma)$. Hence if a  measurement scheme is to lead
to objective pointer values, it will yield no
information at all about the object.

There remains then a last resort to genuinely unsharp
pointers: at present it is not clear if a notion of `unsharp
objectification' (made precise in \cite{uob}), obtained by
dropping (PVD), would allow to evade the measurement
problem. I suspect it would not.  Then one would be forced
to consider one of the options:  giving up the axiom of
linear (unitary) dynamics; considering quantum theories
allowing for (some) classical observables (for measuring
systems); or considering modified notions of the
objectivity, or definiteness, of (pointer) values.  Here I
will consider briefly the last possibility which amounts to
questioning (PM) and pursuing instead some `no-collapse'
interpretation of quantum mechanics.

\section{Modal interpretation for unsharp measurements?}

The question I want to raise in this concluding section is
whether a viable version of modal interpretation could be
developed that would provide an adequate account of what is
happening to an object system in the course of an unsharp
measurement. In view of the fact that virtually any
measurement is imperfect so that the measured observable is
unsharp, it appears to me that in the current versions of
modal interpretations too much emphasis is put on the
idealised case of sharp measurement of discrete observables
and the ensuing biorthogonal decompositions. I would imagine
the following as a promising alternative scenario: It
appears to be the case that some observable of the measuring
system is `naturally' singled out as a suitable pointer
observable whose readings indicate firstly the value of the
measured observable and secondly what can be known about
the object after the measurement. As a matter of fact, experimenters
seem to be bound (by Nature?) to take resort to certain quantities
as indicators of measurement outcomes. Such quantities will on the
one hand have definite values at (practically) all times; but on the
other hand they will be sufficiently sensitive to small-scale
influences so as to be able to probe a microscopic
system. So if the state 
$\rho'_{\cal SA}$ represents the compound system after the
coupling, and if (PM) is {\sl not} stipulated, then 
 -- {\sl given that this state goes along with the
occurrence of a definite pointer reading} -- a simple and
unique prescription for expressing what happened to the
object and apparatus (or probe) would be to assign to them the states 
conditional to a particular pointer reading:
 $$\rho'_{\cal S}(X_i)\ \equiv\
{\rm tr}_{\cal A} \big[\rho'_{\cal SA}(X_i)\big];\ \ 
\rho'_{\cal A}(X_i)\ \equiv\
{\rm tr}_{\cal S} \big[\rho'_{\cal SA}(X_i)\big].  
$$
Here ${\rm tr}_{\cal A}[\cdot]$ denotes the partial trace
with respect to the apparatus Hilbert space, etc.
 The ensuing state pairs $\rho'_{\cal A}(X_i)$ and
$\rho'_{\cal S}(X_i)$ would not in general form a
biorthogonal family; but it seems worthwhile to investigate
whether they can be consistently regarded as the actual
physical states of the probe and object after the
measurement interaction. It may be noted that for sharp
pointers, (PVD) will be automatically satisfied and implies
the mutual orthogonality of the conditional pointer states.
Then a necessary and sufficient condition for the
conditional object states to be orthogonal is the following
{\sl pointer mixture condition for} $\cal A$: 
$$
{\rm tr}_{\cal S}\big[\rho'_{\cal SA}\big] \ =\ 
\sum \rho'_{\cal A}(X_i)
$$
for all $\rho$. This is to say that the reduced apparatus
state assumes the desired form of a mixture of definite
pointer states. See Theorem 3.11 of \cite{cor}. The ``modal rule"
formulated above seems to be quite in line with the
experimenters' actual practice: the said conditional states
do represent the predictive content of the situation reached
after a definite measurement outcome has been recorded. I must leave
it open whether the rule given does amount to a proper interpretation
in the sense of internal consistency.

I will sketch two simple idealised models involving discrete
observables in order to illustrate the state assignment rules
proposed here. This will also make it evident that the 
ensuing kind of modal interpretation that could accommodate
these rules will deviate from the existing variants.

The first model describes an imperfect measurement of a sharp
observable where the imperfection lies in the unitary coupling
not forcing the object system into an eigenstate of the measured
observable. Consider an object described by a 2-dimensional
Hilbert space $\cal H$, and let $\varphi_1,\varphi_2$ be the two
orthogonal normalised eigenvectors of the observable to be
measured. Let $\xi_1,\xi_2$ be two other, nonorthogonal unit vectors
in $\cal H$. Let ${\cal H}_{\cal AE}$ denote the Hilbert
space of the apparatus (including microscopic probes) plus
environment (including observers). We assume the initial state
of $\cal AE$ to be represented by a unit vector
$\phi\in{\cal H}_{\cal AE}$, while the final state
is described in terms of two orthogonal unit vectors
$\phi_1, \phi_2\in{\cal H}_{\cal AE}$. The role of the
environment is to ensure the selection of the pointer states
which in this example are supposed to be mutually orthogonal.
It is not hard to show that the following allows an extension
to a unitary map on the compound Hilbert space: for any initial
state $\varphi=c_1\varphi_1+c_2\varphi_2\in\cal H$, let
$$
\Psi:=\varphi\otimes\phi=\left(c_1\varphi_1+c_2\varphi_2\right)
\otimes\phi\ \longrightarrow\ \Psi':=
c_1\xi_1\otimes\phi_1+c_2\xi_2\otimes\phi_2.
$$
The reading of outcomes would be carried out by looking at
the positions of pointers. This may be modelled by two mutually
orthogonal projections $P_1,P_2$ on ${\cal H}_{\cal AE}$ that are
such that $\left\langle\phi_i|P_k\phi_i\right\rangle=\delta_{ik}$.
Then the measured observable according to the probability
reproducibility condition (PR) is the projection valued measure
with range $P_{\varphi_1},P_{\varphi_2}$. According to the above
rule, the final object state
conditional on a reading $P_k$ is given by $\rho_{{\cal S},k}=
P_{\xi_k}$. It is clear that the biorthogonal decomposition
of $\Psi'$ involves quite different component states than
$\xi_i,\phi_i$. This example shows that even for sharp measurements
there are situations where one would wish to allow state attributions
from a nonorthogonal family of states. It may be noted that
this model is an example of the statement of Theorem III.2.3.1
in \cite{qtm}.

The second example concerns a genuinely unsharp measurement, where
the unsharpness arises from the nonorthogonality of probe states.
The example is modelled after the Stern Gerlach experiment.
The object system will be described as above, but this time the
compound system consisting of probe ${\cal A}$ 
plus environment $\cal E=\cal B+\cal C$ [screen ($\cal B$) 
plus recording device ($\cal C$)]
will explicitly
appear as a compound system, whose initial state will be
taken to be $\phi\otimes\theta$ for simplicity. We assume the
measurement evolution takes place in three stages. Stage 1 is
$$
\Psi:=\left(c_1\varphi_1+c_2\varphi_2\right)\otimes\phi\otimes\theta
\ \longrightarrow\
\Psi':=\left(c_1\varphi_1\otimes\phi_1+c_2\varphi_2\otimes
\phi_2\right)\otimes\theta.
$$
Unitarity is again easy to ensure. This time the probe states
$\phi_1,\phi_2$ are assumed to be nonorthogonal. In the Stern
Gerlach experiment, these states correspond to the spin-$\frac 12$
particle's spatial wavefunctions correlated with the spin
eigenstates $\varphi_1,\varphi_2$, and due to wave packet spreading
they will have a spatial overlap and generally nonvanishing inner
product. Stage 2 consists in the production of a scintillation
in the upper or lower screen half, thus indicating which way the
particle appears to have taken; we write this step for the two
components of $\Psi'$ separately:
\begin{eqnarray*}
\varphi_1\otimes\phi_1\otimes\theta\ &\longrightarrow\ \Psi_1'':=
\varphi_1\otimes\left(\alpha_{11}\phi_{11}\otimes\theta_{11}+
\alpha_{12}\phi_{12}\otimes\theta_{12}\right),\\
\varphi_2\otimes\phi_2\otimes\theta\ &\longrightarrow\ \Psi_2'':=
\varphi_2\otimes\left(\alpha_{21}\phi_{21}\otimes\theta_{21}+
\alpha_{22}\phi_{22}\otimes\theta_{22}\right),\\
&\Psi'':=c_1\Psi_1''+c_2\Psi_2''.
\end{eqnarray*}
Here $\theta_{11},\theta_{21}$ represent states of the environment
corresponding to sensitive molecules localised in the upper
half of the screen, while $\theta_{22},\theta_{12}$ correspond to
molecules localised in the lower half. Thus, if the wave packet
$\phi_1$ [$\phi_2$] were strictly localised in the upper [lower]
screen region then only the component containing $\theta_{11}$
[$\theta_{22}$] would emerge. But realistically, the tail of
$\phi_1$ evolving to the lower screen half allows for a chance
of the particle to interact with molecules of that ``wrong"
section. We assume the $\theta_{ij}$ to be normalised and mutually
orthogonal:
$$
\left\langle\theta_{ik}|\theta_{jl}\right\rangle=
\delta_{ij}\delta_{kl}.
$$
 Further, we require $\alpha_{ij}>0$, $\alpha_{11}^2+
\alpha_{12}^2=1$, $\alpha_{21}^2+\alpha_{22}^2=1$.
Then a unitary extension of this map is guaranteed to
exist. We also assume that the vectors $\phi_{ij}$ are
normalised and mutually orthogonal, representing wave packets
localised in the appropriate screen halves.

The third stage, the recording, is modelled by taking into account
that the states $\theta,\theta_{ij}$ are product states of the form
$$
\theta=\chi\otimes\psi,\quad
\theta_{ij}=\chi_{ij}\otimes\psi,
$$
where the first factor represents the states of the sensitive
screen molecule, while the second factor is the state of the
recording device (observer). Then after the completion of the
second stage, the recording takes place in such a way that
the states $\theta_{ij}$ evolve independently of the object
plus probe:
$$
\theta_{ij}\longrightarrow\chi_{ij}\otimes\psi_j,
$$
where $\psi_1$ [$\psi_2$] represents the state that the recording device
(observer) has reached after registering a scintillation in the upper
[lower] screen half. These states are assumed to be orthogonal.
Hence,
\begin{eqnarray*}
\Psi''\ \longrightarrow\ \Psi'''\ :=\ 
\left[c_1\varphi_1\otimes\alpha_{11}\phi_{11}\otimes\chi_{11}
+c_2\varphi_2\otimes\alpha_{21}\phi_{21}\otimes\chi_{21}\right]
\otimes{\psi_1}\\
+\left[c_1\varphi_1\otimes\alpha_{12}\phi_{12}\otimes\chi_{12}
+c_2\varphi_2\otimes\alpha_{22}\phi_{22}\otimes\chi_{22}\right]
\otimes{\psi_2}\\
\end{eqnarray*}
This is conveniently described by projections $P_1,P_2$
acting in the screen Hilbert space,
indicating whether a scintillation has occurred on the upper or lower
screen half. Hence,
$$
\langle\theta_{11}|P_1\theta_{11}\rangle=1,\quad
\langle\theta_{21}|P_1\theta_{21}\rangle=1,\quad
\langle\theta_{12}|P_2\theta_{12}\rangle=1,\quad
\langle\theta_{22}|P_2\theta_{22}\rangle=1.
$$
Given that the actual state of the observer is one of $\psi_1,\psi_2$,
the conditional final states of object and probe are given by the
following equations:
$$
\left\langle \Psi'''|X\otimes I_{\cal A}\otimes I_{\cal B}\otimes
P_{\psi_i}\,\Psi'''\right\rangle\ =\
\left\langle\Psi''|X\otimes I_{\cal A}\otimes P_i|,\Psi''\right\rangle\ =:\
{\rm tr}[\rho_{{\cal S},i}X],
$$
$$
\left\langle \Psi'''|I_{\cal S}\otimes Y\otimes I_{\cal B}\otimes
P_{\psi_i}\,\Psi'''\right\rangle\ =:\ {\rm tr}[\rho_{{\cal A},i}Y].
$$
Here $X,Y$ are arbitrary bounded selfadjoint operators on the object
and probe Hilbert spaces, respectively. It follows that the actual
states of $\cal S$ and $\cal A$ conditional on a reading $i$ are
\begin{eqnarray*}
\rho_{{\cal S},i}\ =\ \alpha_{1i}^2P_{\varphi_1}P_{\varphi}P_{\varphi_1}
+\alpha_{2i}^2P_{\varphi_2}P_{\varphi}P_{\varphi_2},\\
\rho_{{\cal A},i}\ =\ |c_1|^2\alpha_{1i}^2P_{\phi_1i}+
|c_2|^2\alpha_{2i}^2P_{\phi_2i}.\\
\end{eqnarray*}
Notice that one would like to have $\alpha_{11}^2\approx 1
\approx\alpha_{22}^2$
and $\alpha_{12}^2\approx 0\approx\alpha_{21}^2$,
so that $\cal S$ would be nearly
in one of the states $\varphi_i$ and $\cal A$ nearly in the corresponding
state $\phi_i$. But as it stands, both systems are in states which are
proper quantum mixtures, that is, they do not allow an ignorance
interpretation with respect to their pure components. This is due to
the fact that according to our state ascription rule the system
$\cal S+\cal A+\cal B$ is in one of the pure states strictly correlated with
the states $\psi_i$ of the registration device; and in the situation
modelled here there is ``nothing in the world" that could allow one to
perform a pure state ascription to subsystems in these pure entangled states.

Finally we note that the observable measured in this model is given by
$$
\left\langle\Psi'''|I_{\cal S}\otimes I_{\cal A}
\otimes I_{\cal B}\otimes P_{\psi_i}
\Psi'''\right\rangle\ =\
\left\langle\Psi''|I_{\cal S}\otimes I_{\cal A}\otimes P_i
\otimes I_{\cal C}\Psi'' \right\rangle\
=:\
\langle\varphi|E_i\varphi\rangle.
$$
This gives the following effect operators:
$$
E_i\ =\ \alpha_{1i}^2P_{\varphi_1}+\alpha_{2i}^2P_{\varphi_2}.
$$ 
Under the above assumptions on the $\alpha_{ij}$ 
the $E_i$ would be nearly equal to $P_{\varphi_i}$. Notice that
then $E_i$ as well as $P_{\varphi_i}$ have high probabilities
to occur in the object state $\rho_{{\cal S},i}$;  this is to say
that operationally these positive operators correspond to some elements of
{\sl unsharp reality} \cite{oqp}.

These examples make it clear that the state ascription rule applied here
relies on a hierarchical order of the observables: there need to be
mechanisms by which some observables of macroscopic systems are
selected by Nature to have definite values throughout and to be able to
correlate with observables of microsystems so as to monitor their values
and make them inherit their own definiteness. The case of a closed
system is included as a limiting case: if an object system does not interact
with the given macroscopic device or observer, its reduced state will
always be the state that evolved from its original state due to its
free evolution.

As a perhaps somewhat more realistic example of a measurement
in which the final object states
are not mutually orthogonal, one may think of a `preparatory'
phase space
measurement in which the said reduced conditional object states
correspond to situations in which the system is fairly well
though only unsharply localised in some phase space cell
indicated by the reading. This is to say that the particle
in question has a strong tendency (high probability) to
`show up' in that region in a subsequent phase space
measurement. 

The quest for such a variation of a modal interpretation
becomes even more urgent if one accepts the idea that
macroscopic pointers are intrinsically unsharp observables
themselves: in that case the pointer states 
are not mutually
orthogonal either. [Notice that in the second example
above the final probe states $\rho_{{\cal A},i}$ are not mutually
orthogonal either.] One may  model such a
situation by describing the pointer again as a phase space
observable -- after all, pointers are objects localised in
space and with fairly definite and controllable speeds.
These quasi-classical features require unsharpnesses in
positions and momenta that are huge compared to the scale
given by Planck's constant.

The kind of nonorthogonal state attribution, along with the
unsharp value-attribution proposed here surely
needs to be analysed in regard to its internal
consistency. Also, one may ask by what mechanisms (if any)
some specific unsharp macroscopic observables should be
selected by Nature and enabled to serve as pointers. On the
other hand it appears to me that the unsharpness involved
could offer new flexibility in dealing with some of the
difficulties encountered with current modal
interpretations. For example, collections of unsharp
observables can be coexistent -- i.e., have joint
distributions -- even if they do no commute. This may help
to avoid Kochen-Specker-type inconsistencies arising from
simultaneous value assignments to noncommuting
quantities. And this could also have implications on the
notorious degeneracies problem since it would be conceivable
that the simultaneous (quasi-)diagonality of a mixed state
in different bases corresponding to noncommuting quantities
is consistently interpretable as all these quantities having
definite (unsharp) values.


\begin{thebibliography}{99}
\bibitem{oqp} P.  Busch, M. Grabowski, P. Lahti (1995).
{\sl Operational Quantum Physics}, Springer-Verlag,
LNP Vol. m31, Berlin. 
\bibitem{neu} J. von Neumann (1955). {\sl Mathematical Foundations
of Quantum Mechanics}, Princeton University Press,
Princeton, Sec.\ V.4.
\bibitem{lud} G. Ludwig (1987). {\sl An Axiomatic Basis for Quantum
Mechanics, Vol 2: Quantum Mechanics and Macrosystems}.
Springer, Berlin.
\bibitem{insol} P. Busch, A. Shimony (1996). Insolubility of
the Quantum Measurement Problem for Unsharp Observables,
{\sl Stud. Hist. Phil. Mod. Phys. 27} 397-404. 
\bibitem{stein} H. Stein (1997). On a maximal impossibility theorem
for the quantum theory of measurement. In: {\sl Potentiality,
Entanglement, and
Passion at a Distance: Quantum Mechanical
Studies for Abner Shimony}, eds. R.S. Cohen, M.A. Horne, J.
Stachel, Kluwer, Dordrecht, in press.
\bibitem{uob} P. Busch (1997). Can `Unsharp Objectification'
Solve the Quantum Measurement Problem? {\sl Int. J. Theor.
Phys. 36}, to appear.
\bibitem{qtm} P. Busch, P. Lahti, P. Mittelstaedt (1996). 
{\sl The Quantum Theory of Measurement}, Springer, LNP Vol.
m2, 2nd revised edition.
\bibitem{cor} P. Busch, P. Lahti (1996). Correlation
properties of quantum measurements, {\sl J. Math. Phys. 37}
2585-2601. 
\end{thebibliography}
\end{document}